\documentclass[prl,aps,nobalancelastpage,amssymb,twocolumn]{revtex4}

\usepackage{graphicx}

\def \beqn {\begin{equation}}
\def \bfig {\begin{figure}}
\def \btab {\begin{table}}
\def \eeqn {\end{equation}}
\def \efig {\end{figure}}
\def \etab {\end{table}}

\begin{document}

\title{Irradiation-induced confinement in a quasi-one-dimensional metal}

\author{A. Enayati-Rad$^1$, A. Narduzzo$^1$, F. Rullier-Albenque$^2$, S. Horii$^3$ \& N. E. Hussey$^1$}

\affiliation{$^1$H. H. Wills Physics Laboratory, University of Bristol, Tyndall Avenue, BS8 1TL, U.K.}

\affiliation{$^2$SPEC, Orme des Merisiers, CEA, 91191 Gif sur Yvette cedex, France}

\affiliation{$^3$Department of Applied Chemistry, University of Tokyo, 7-3-1 Hongo, Tokyo 113-8656, Japan}

\date{\today}

\begin{abstract}
The anisotropic resistivity of PrBa$_2$Cu$_4$O$_8$ has been measured as a function of electron irradiation fluence.
Localization effects are observed for extremely small amounts of disorder corresponding to electron mean-free-paths of
order 100 unit cells. Estimates of the localization corrections suggest that this anomalous localization threshold
heralds a crossover to a ground state with pronounced one-dimensional character in which conduction electrons become
confined to a small cluster of chains.
\end{abstract}

\maketitle

Disorder-induced metal/insulator transitions (MIT) have been a dominant theme in solid state research for several
decades yet many outstanding issues remain. In three-dimensional (3D) metals,  metallic behavior is observed only for
$k_F \ell_0$ (the product of the Fermi wave vector and the mean-free-path) $>$ 1 \cite{Mottbook}. The insulating state,
characterized by a vanishing of $\sigma_0$ the dc conductivity at $T$ = 0, occurs once the carrier concentration
($\propto k_F$), disorder potential ($\propto 1/\ell_0$) or some other control parameter reach a critical value.
Neither this nor the critical exponents at the transition however, are known precisely \cite{Gantmakher}. Scaling
theory advocates no genuine metallic state in two dimensions \cite{Abrahams79} yet in 2D electron gases, a MIT is
observed at the universal sheet resistance 2$h/e^2$ \cite{Kravchenko95}.

In strictly 1D systems, all electronic states are localized at $T$=0 in the presence of weak disorder \cite{MottTwose}.
In real materials with finite interchain coupling $t_{\bot}$ however, the situation is not so clear. Abrikosov and
Ryzchkin claim that {\it any} finite $t_{\bot}$ stabilizes the metallic state \cite{AbrikosovRyzhkin} whilst Prigodin
and Firsov (PF) argue that impurity scattering rates $\hbar/\tau_0 > t_{\bot}$ render the system effectively 1D and
therefore susceptible to localization at low $T$ \cite{PrigodinFirsov}. This controversy has never been resolved
experimentally, due to the lack of {\it quantitative} studies of the MIT in quasi-1D conductors and the lack of systems
of the appropriate dimensionality displaying low-$T$ metallic behavior.

Irradiation experiments are a reliable means of controllably changing the defect density in solids with the added
advantage that transport properties can be measured {\it in-situ}. In this Letter, we report a systematic study of the
effects of electron irradiation on the anisotropic resistivity of the quasi-1D cuprate PrBa$_2$Cu$_4$O$_8$ (Pr124).
Pr124 comprises a 1D network of zig-zag double chains (oriented along the $b$-axis) sandwiched between insulating
CuO$_2$ bilayers (in the $ab$-plane). In clean Pr124, a highly anisotropic but nevertheless 3D metallic state develops
at low $T$ with $\rho$($T$) $\sim$ $T^2$ for ($\rho_a$:$\rho_b$:$\rho_c$ $\sim$ 1000:1:3000) \cite{McBrien02,
Hussey02}. With increasing irradiation, low-$T$ upturns in $\rho$($T$) develop {\it simultaneously} for {\bf I}$\| a,b$
and $c$ once $k_F \ell_0 <$ 60. For {\bf I}$\| b$ (in-chain conductivity), these upturns are consistent with 1D
inter-electron interference corrections of a magnitude that implies confinement of the charge carriers to a very small
number of chains.

\begin{figure}
\includegraphics[width=7.0cm,keepaspectratio=true]{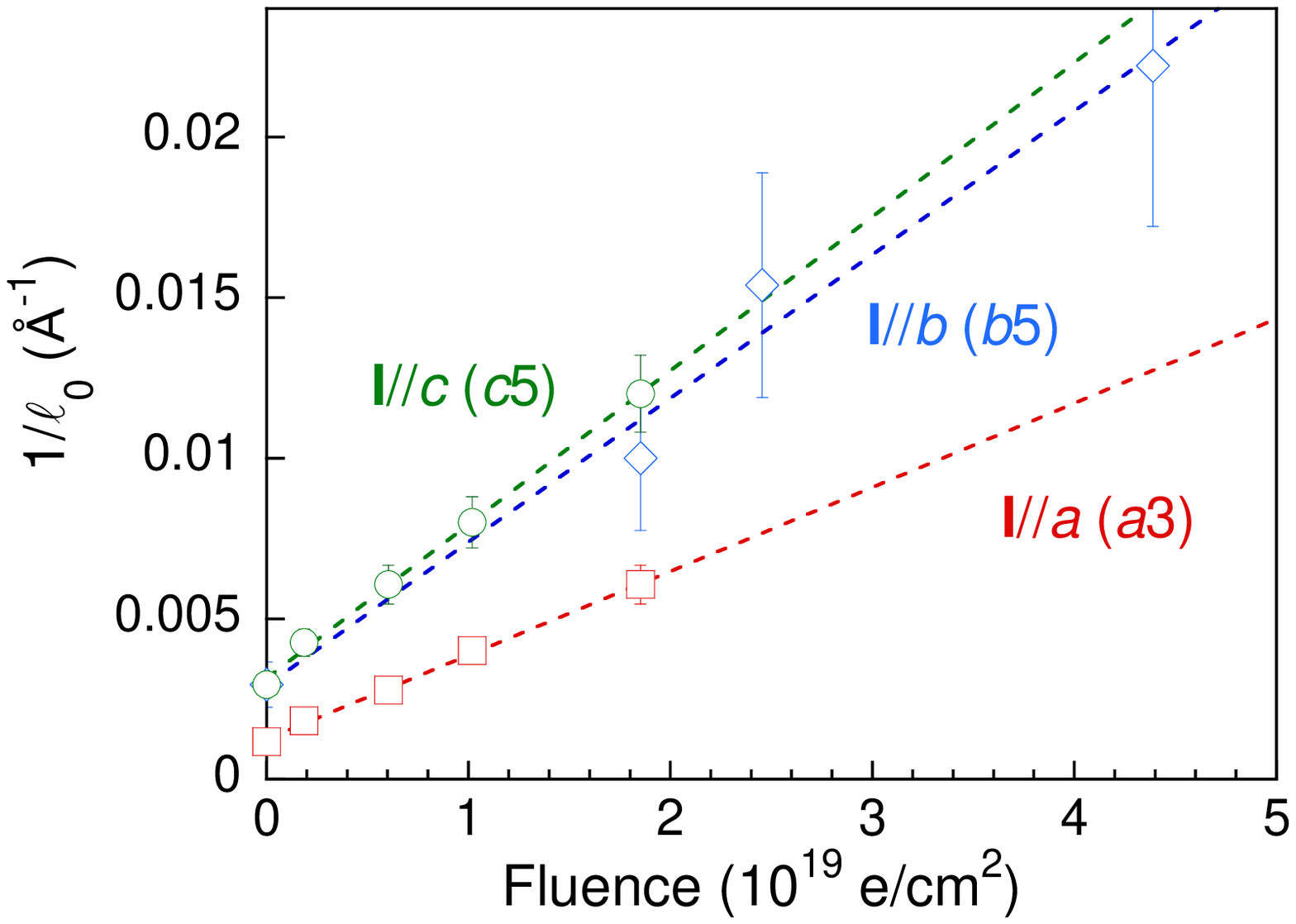}
\caption{(Color online) $1/\ell_0$ vs. fluence $\phi$ for three Pr124 crystals $a3$, $b5$ and $c5$ with {\bf I}$\|a$,
$b$ and $c$ respectively. Error bars for {\bf I}$\|b$ are due to uncertainties in the crystal dimensions. For {\bf
I}$\|a$ and {\bf I}$\|c$, error bars are estimated from a relative measurement (see text for details) and are thus
significantly smaller.} \label{fig1}
\end{figure}

A set of crystals (see Ref.\cite{Horii00} for growth details) of appropriate geometries were mounted in different
contact configurations to ensure uniaxial current flow in one of the three orthogonal directions. For the irradiation
experiments (performed at Ecole Polytechnique, Palaiseau, France) samples were cooled in liquid hydrogen to 20K,
irradiated in a 2.5MeV electron beam, taken to 150K then cycled back to 20K. $\rho(T)$ was measured simultaneously
using a standard four-probe dc technique. The thicknesses of the samples were small compared to the penetration depth
of the electrons ensuring a homogeneous damage throughout. After the irradiation experiments, some further $\rho$($T$)
measurements were performed (on those crystals whose electrical contacts had survived) in a different (ac) setup down
to 4.2K.

\begin{figure*}
\includegraphics[width=7.0cm,angle=270,keepaspectratio=true]{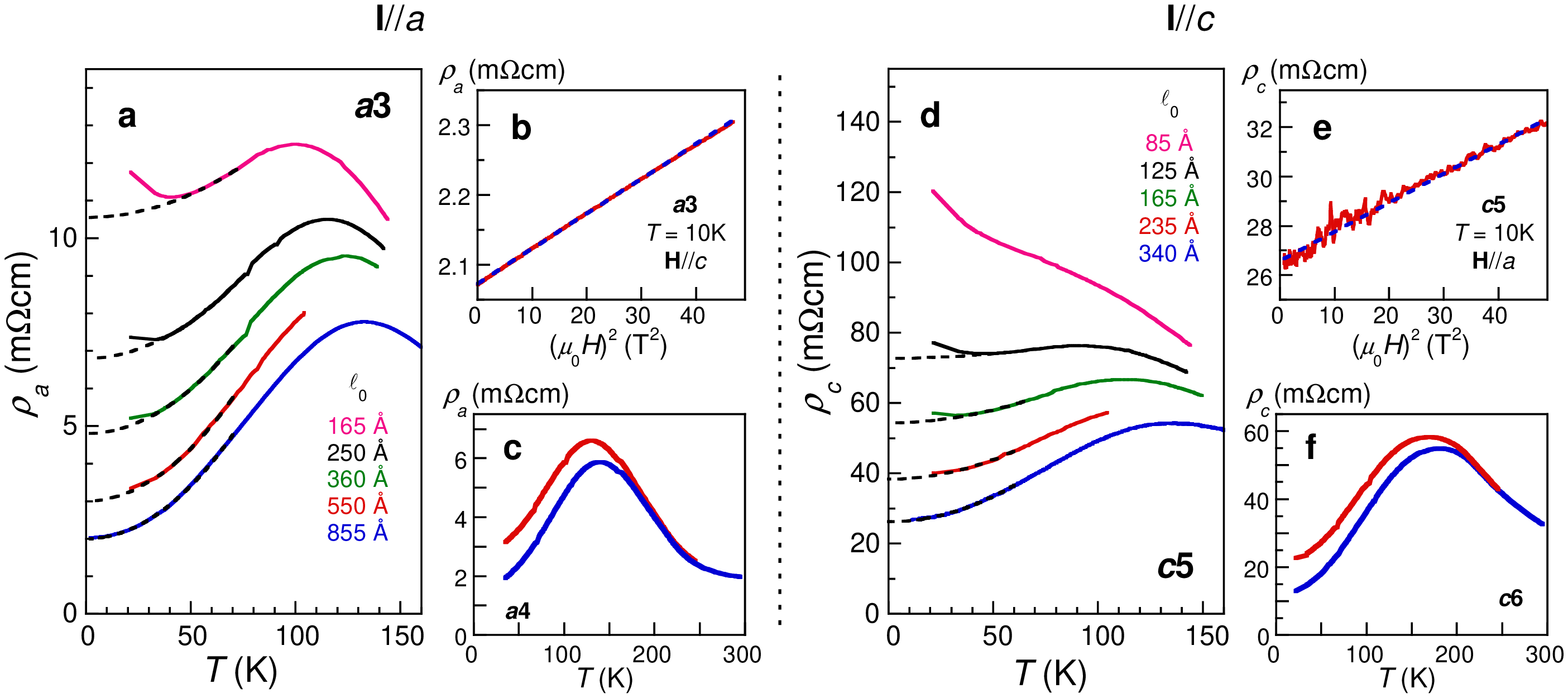}
\caption{(Color online) a) $\rho_a$($T$) of $a3$ for different irradiation fluences. Dashed lines are extrapolations of
the $T^2$ resistivity down to 0K. $\ell_0$ values for each fluence (see text for details) are also listed. b) $\rho_a$
vs. $H^2$ for $a3$ at $T$ = 10K and {\bf H}$\|c$. c) $\rho_a$($T$) of a second crystal ($a4$) before (lower) and after
(upper curve) irradiation. d) $\rho_c$($T$) of $c5$ subjected to the same fluences as $a3$. e) $\rho_c$ vs. $H^2$ for
$c5$ at $T$ = 10K and {\bf H}$\|a$. f) $\rho_c$($T$) of a second crystal ($c6$) before (lower) and after (upper curve)
irradiation.} \label{fig2}
\end{figure*}

For quantitative analysis, reliable estimates of $\ell_0$ are required at each stage of the experiment. For {\bf
I}$\|b$, $\ell_0$ is obtained from the residual resistivity $\rho_0$ using the Drude formula for a (double-chain) 1D
metal $\ell_0$ = $\hbar ac/4\pi e^2\rho_0$ where $a$ = 3.86$\AA$ and $c$ = 13.6$\AA$ are lattice constants. For
interchain currents, $\ell_0$ is determined from low-field transverse magnetoresistance (MR) measurements carried out
on the non-irradiated (virgin) crystals at $T$ = 10K. For {\bf I}$\|a$, Boltzmann transport theory predicts $\ell_0$ =
$\hbar/ea$($\Delta\rho_a/\rho_a)^{0.5}$ where $\Delta \rho_a/\rho_a$ = $\Delta \rho_a(\mu_0 H$=1T)/$\rho_a(\mu_0 H$=0)
for {\bf H}$\|c$ \cite{Horii02}. (For {\bf I}$\|c$ and {\bf H}$\|a$, one simply substitutes $a$ with $c$.) The $\ell_0$
values of irradiated crystals are obtained by scaling each (extrapolated) $\rho_0$($\phi$) to the corresponding
$\rho_0$ of the virgin crystal(s). Fig. \ref{fig1} shows a plot of 1/$\ell_0$ vs. fluence $\phi$ for the three samples
of main interest in this study. The linear scaling of 1/$\ell_0$ for each crystal confirms the reliability of the
extrapolations (dashed lines in Fig. 2a, 2d and 3). Moreover, the shifts in 1/$\ell_0$ for {\bf I}$\|b$ and {\bf
I}$\|c$ are the same within experimental uncertainty suggesting that the estimates for $\ell_0$ in these two cases are
robust whilst those for {\bf I}$\|a$ are overestimated by $\sim 60\%$.

The effects of irradiation on the two interchain resistivities $\rho_a$ and $\rho_c$ are compared in Fig. \ref{fig2}.
Fig. 2a and 2d show sets of $\rho_a$($T$) and $\rho_c$($T$) curves for crystals $a3$ and $c5$ respectively. (The MR
sweeps from which the virgin $\ell_0$ values were obtained are shown in Fig. 2b and 2e.) Prior to irradiation (bottom
curves in Fig. 2a and 2d), $\rho_{a(c)}$($T$) is metallic below $T_{\rm max} \sim$ 150K and varies quadratically with
temperature below $\sim$ 70K. Although Matthiessen's rule is clearly violated in both cases, the fact that 1/$\ell_0$
scales linearly with fluence (Fig. \ref{fig1}) implies that the impurity scattering term remains additive. The
violation of Matthiessen's rule however does suggest a gradual crossover from coherent to incoherent interchain
hopping, the peaks in $\rho_{a(c)}$($T$) at $T$ = $T_{\rm max}$ gradually shifting to lower $T$ as the fluence is
increased. Above $T_{\rm max}$, $\rho_{a(c)}$($T$) is largely unaffected by the irradiation, as highlighted in Fig. 2c
and 2f for two further crystals $a4$ and $c6$. This insensitivity to disorder favors phonon-assisted rather than
impurity-assisted hopping in Pr124 at high $T$ \cite{Analytis06}. For the largest fluences, an upturn in the
resistivity is observed at $T$ = $T_{\rm min}$. Inspection of Fig. 2a and 2d reveals that both $T_{\rm min}$ and
$T_{\rm max}$ are comparable for similar values of $\ell_0$, indicating that the evolution of $\rho_a$($T$) and
$\rho_c$($T$) with disorder is essentially the same. Given that for $\ell_0 < 100\AA$ (top curve in Fig. 2d), the peak
in $\rho_c$($T$) vanishes altogether, we conjecture that for $\ell_0 < 100\AA$, both $\rho_a$ and $\rho_c$ are
incoherent.

\begin{figure}
\includegraphics[width=6.5cm,keepaspectratio=true]{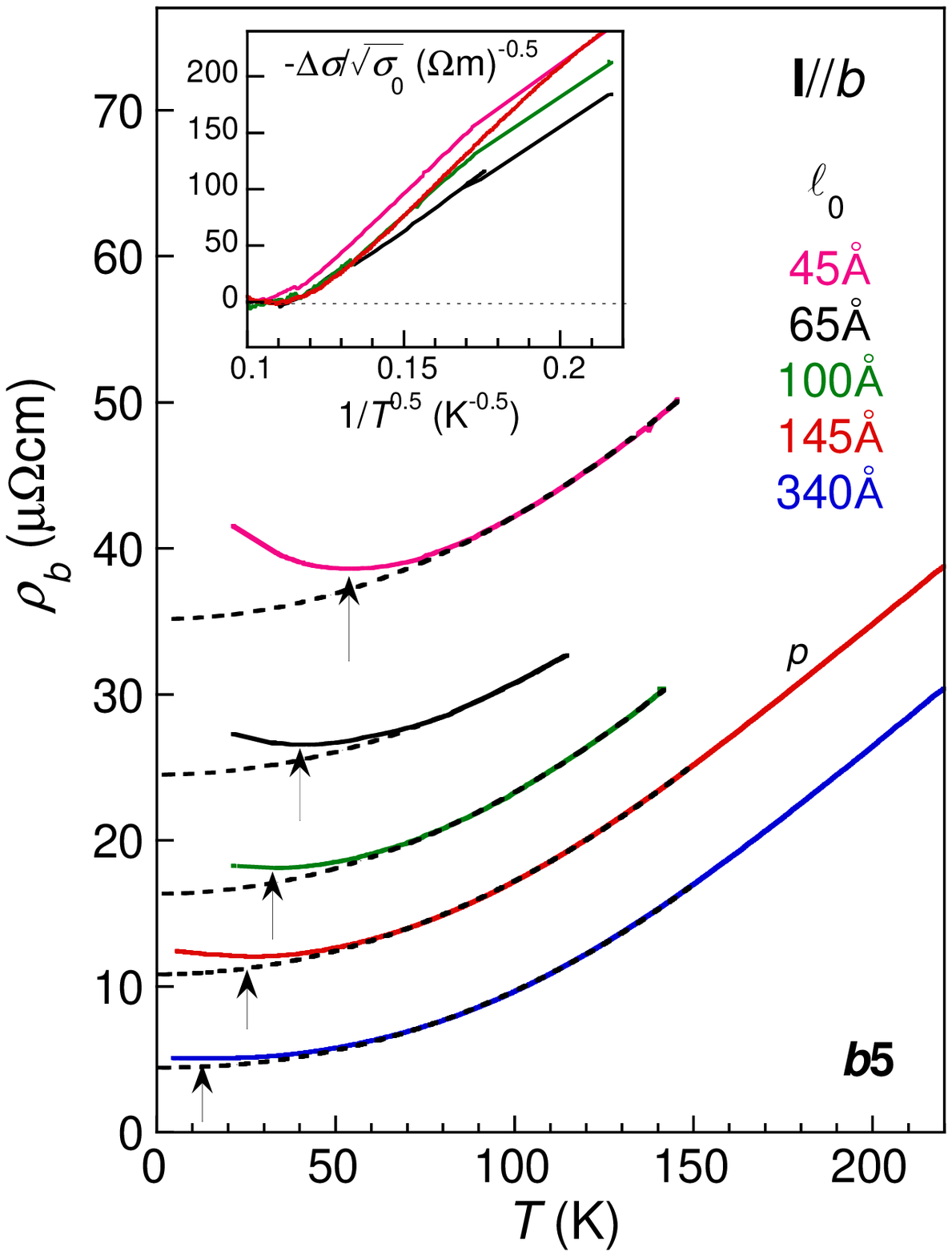}
\caption{(Color online) $\rho_b$($T$) of Pr124 (crystal $b5$) for different fluences with the corresponding $\ell_0$
values listed in the top right-hand corner. Arrows indicate $T_{\rm min}$ and dashed lines are extrapolations of the
$T^2$ resistivity down to 0K. $p$ stands for post-irradiation measurement. Inset: - $\Delta\sigma/\surd\sigma_0$ vs.
1/$T^{0.5}$.} \label{fig3}
\end{figure}

Let us now turn to the in-chain current response {\bf I}$\|b$. Fig. \ref{fig3} shows $\rho_b$($T$) for crystal $b5$
(subjected to higher fluences than $a3$ and $c5$). Again, the arrows indicate $T_{\rm min}$ for each fluence and the
corresponding $\ell_0$($\phi$) values, obtained from the extrapolation of the $T^2$ resistivity to 0K (dashed lines in
Fig. \ref{fig3}), are appropriately color-coded. In contrast to the interchain response, Matthiessen's rule clearly
holds for {\bf I}$\|b$ affirming that the in-chain carrier density is unaffected by electron irradiation and that the
interchain current response is indeed governed by the coherent-incoherent crossover. In addition to the upward shift in
$\rho_b$($T$), a resistive upturn, barely visible in the virgin curve, begins to develop. The curve labelled $p$,
measured {\it after} the crystal was removed from the irradiation probe, demonstrates that upon warming, much of the
irradiation damage is annealed out, presumably due to recombination processes of close vacancy-interstitial pairs and
migration of interstitial atoms \cite{Rullier01}.

In order to gain further understanding of the nature of the upturns in sample $b5$, we examined the correction term
$\Delta\sigma(T)$ = $1/\rho_b(T) - \sigma_{\rm D}(T)$ where $\sigma_{\rm D}(T) = 1/(A+BT^2)$ are the dashed curves in
Fig. 3. Although our useful fitting range was somewhat limited, the data did not appear to follow the $T$-dependence
expected for 1D weak localization (WL) corrections ($\propto T^{-p/2}$ where $p$ = 2 is the exponent of the $T$
variation of the inelastic scattering time \cite{Gantmakher}), Kondo scattering ($\propto $ ln$T$ \cite{Kondo}) nor WL
corrections for a Luttinger liquid ($\propto T$ln$T$ \cite{Gornyi05}). Instead the data for 20K $\leq T \leq$ 100K
could best be described by the form expected for 1D Altshuler-Aronov (AA) inter-electron interference corrections
\cite{AltshulerAronov}, $\Delta\sigma = (2-3F/2)(e^2/\pi \hbar A) (\hbar D/k_BT)^{1/2}$ where $A$ is the
cross-sectional area of the conducting element, $F$ is a screening parameter and $D = v_F\ell_0$ is the diffusion
constant. The Fermi velocity within the Pr124 chains $v_F$ = 2.5 $\times$ 10$^5$ ms$^{-1}$ \cite{Mizokawa00}. Thus
expressing $\ell_0$($\phi$) as a function of $\rho_0$($\phi$) gives $\Delta\sigma/\surd\sigma_0(\phi) = mT^{-1/2} =
1400(2-3F/2)/AT^{1/2}$ for $A$ in nm$^2$ and $\sigma$ in ($\Omega$m)$^{-1}$. As shown in the inset to Fig. \ref{fig3},
a quasi-linear dependence is observed for all irradiations with $m \sim 2000 \pm 200$ (K/$\Omega$m)$^{0.5}$,
corresponding to $A/ac \sim$ 1-3 for 0 $\leq F \leq$ 1. This result confirms not only the presence of strong electron
correlations in Pr124, more importantly it implies that {\it charge transfer in the localized regime is confined to a
very small number of CuO chains}.

\begin{figure}
\includegraphics[width=7.5cm,keepaspectratio=true]{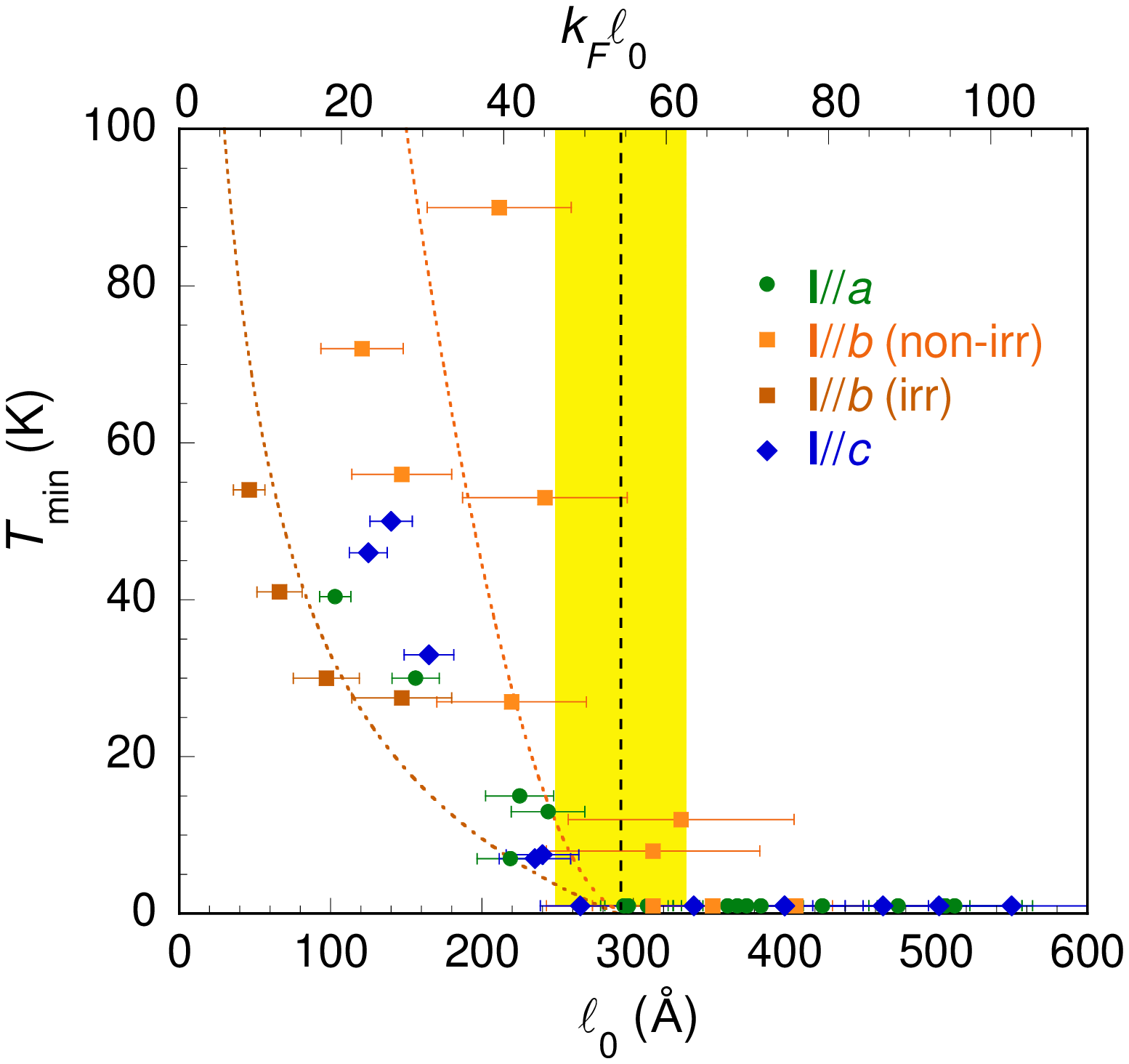}
\caption{(Color online) $T_{\rm min}$ versus $\ell_0$ (bottom) and $k_F\ell_0$ (top abscissa) for both irradiated and
non-irradiated Pr124. The circles, squares and diamonds correspond to {\bf I}$\|a$,$b$,$c$ respectively. The $\ell_0$
values are determined as described in the text, with those for {\bf I}$\|a$ scaled by a factor 1.6 as suggested by Fig.
\ref{fig1}. The shaded region (marked by a dashed line) represents the threshold in $\ell_0$ (and $k_F\ell_0$) beyond
which no resistive upturns are observed. The upper and lower dotted lines delineate non-irradiated from irradiated
$b$-axis crystals respectively.} \label{fig4}
\end{figure}

A compilation of data from $\sim 40$ non-irradiated and irradiated crystals is summarized in Fig. \ref{fig4} which
shows $T_{\rm min}$ plotted versus $\ell_0$ (lower abscissa, determined in the manner described above) and $k_F\ell_0$
(upper abscissa). Here $k_F$ = 0.23$\pi$/$a$ = 0.19$\AA^{-1}$ as obtained from photoemission studies \cite{Mizokawa00}.
The points lying along $y$ = 0 represent those resistivity curves that exhibited no upturn down to the lowest
temperatures measured (between 0.4K and 5K). Whilst $T_{\rm min}$ itself has little physical significance, the plot
clearly defines a threshold, indicated by the shaded region, beyond which no upturn in $\rho$($T$) is observed,
implying that the localization threshold is a fundamental property of Pr124.

To the left of the shaded region, the data for {\bf I}$\parallel$$b$ show considerable scatter, due to uncertainty in
the sample dimensions, but also appear to bunch into two distinct groups of crystals; irradiated (lower) and
non-irradiated (upper dotted line) respectively. This delineation is intriguing and warrants further explanation. Mg,
the principal defect in our crystals, substitutes preferentially onto the chains due to its lack of pyramidal
co-ordination. This is supported by electron probe microanalysis that indicates a close correspondence between $\ell_0$
and the distance between Mg atoms \cite{Narduzzo06}. Electron irradiation, by contrast, creates defects randomly. In
YBa$_2$Cu$_3$O$_{7-\delta}$, for instance, irradiation suppresses superconductivity in a quantitatively similar manner
to Zn substitution, without changing the doping level \cite{Legris93}, implying a significant defect density within the
CuO$_2$ planes. Once localization develops, in-chain defects will induce the more pronounced localization effects due
to the predominance of large-angle scattering. Hence in crystals with high Mg content, $T_{\rm min}$ is enhanced. For
{\bf I}$\parallel$$a,c$, the data points lie between these two extremes and are insensitive to the type of disorder, as
one might expect since these current paths involve both chain and planar sites.

Irrespective of these details, the threshold itself appears to be independent of the direction of current flow,
affirming that the onset of localization occurs {\it simultaneously} along all three axes and is determined by the {\it
total} (small- and large-angle) scattering rate for the in-chain carriers. The shaded region in Fig.~\ref{fig4}
corresponds to $\ell_0 \sim 300 \pm 50 \AA$ or equivalently $k_F\ell_0 \sim 55 \pm 10$. Whilst this seems a large value
for pronounced localization effects to be observed, quantum interference and correlation effects are known to be
enhanced in disordered systems of reduced dimensionality. Indeed in Pr124, $t_{\bot}$ is extremely small ($\sim$
2-3meV) in both orthogonal directions \cite{Narduzzo06}. With increasing disorder, quantum corrections thus emerge,
impeding Bloch wave propagation throughout the crystal(s), even though $k_F\ell_0 \sim$ 60 and the original Bloch
states are extended over 10$^2$ unit cells. At the localization threshold, $\hbar/\tau_0 = \hbar v_F/\ell_0 \sim$
3-4meV $\geq t_{\bot}$. According to PF \cite{PrigodinFirsov}, localization in quasi-1D conductors should always occur
for $\hbar/\tau_0 > 3t_{\bot}$; once the intrachain scattering rate surpasses the interchain hopping rate, coherent
interchain tunnelling is blocked, the system is rendered effectively 1D and localization becomes inevitable.

Whilst our data are in good quantitative agreement with the predictions of PF for disorder-induced one-dimensionality,
the dimensional crossover in Pr124 may not be as complete as originally envisaged. The magnitude of the AA corrections
is consistent with carrier confinement to a single chain. However, without {\it a priori} knowledge of the screening
parameter $F$, we cannot rule out carrier motion that extends over a small but finite number of chain units. Indeed, a
recent transport study of non-irradiated Pr124 showed evidence for the presence of an orbital MR within the localized
region \cite{Narduzzo07}. This was interpreted as a coexistence of metallicity on a microscopic length scale (i.e.
metallic chain segments between impurity states) and localization on a global scale, over the length of the sample.
This picture bears certain similarity to the so-called \lq interrupted strand' model (ISM) \cite{KuseZeller,
RiceBernasconi}, that was applied to explain the activated behavior of the dc conductivity in irradiated quasi-1D
organic Peierls insulators \cite{EroGecs79, Zuppiroli80a, Forro82}. In the ISM, defects are assumed to be perfectly
insulating (in the strictly 1D sense) with no coherent tunneling between the metallic islands. In Pr124 by contrast,
the preservation of $t_{\bot}$ and {\it occasional} coherent tunnelling between chains appears crucial for the
existence of the orbital MR \cite{Narduzzo07}. At smaller values of $k_F\ell_0$ (larger values of $\hbar/\tau_0$),
interchain transport does eventually become totally incoherent. This occurs in the region where $k_F\ell_0 \sim$ 15-20
or conversely, when $\hbar/\tau_0 > 4t_{\bot}$, the total interchain bandwidth. It remains to be seen however whether
this coherent-incoherent crossover ultimately drives the system across the MIT.

In summary we have demonstrated the onset of localization in the quasi-1D metal Pr124 through irradiation and intrinsic
disorder. The localization threshold is found to occur simultaneously along all three crystallographic axes once
$k_F\ell_0 <$ 60, almost two orders of magnitude larger than in isotropic 3D metals. This threshold appears consistent
with theoretical predictions for a disorder-induced 3D-1D transition, though certain aspects of the data suggest the
preservation of finite interchain coupling, if only over microscopic length scales, in the regime where $\hbar/\tau_0 >
t_{\bot}$.

We thank A. Carrington and N. Shannon for helpful discussions. This work was supported by the EPSRC.



\end{document}